# A universal framework for nonlinear frequency combs under electro-optic modulation

Yanyun Xue[1,*], Xianpeng Lv[1,*], Guangxing Wu[2,*], Tianqi Lei[1,*], Chenyang Cao[1], Yiming Lei[1], Min Wang[1], Zhendong Zhu[1], Yan Li[1,3,4], Qihuang Gong[1,3,4], Di Zhu[2,†], Yaowen Hu[1,3,†]

[1]State Key Laboratory for Mesoscopic Physics and Frontiers Science Center for Nano-optoelectronics, School of Physics, Peking University, Beijing 100871, China

[2]Department of Materials Science and Engineering, National University of Singapore, Singapore 117575, Singapore

[3]Collaborative Innovation Center of Extreme Optics, Shanxi University, Taiyuan, China.

[4]Hefei National Laboratory; Hefei 230088, China.

[*]These authors contributed equally.

[†]Corresponding authors: dizhu@nus.edu.sg; yaowenhu@pku.edu.cn;

**Nonlinear frequency combs, including electro-optic and Kerr combs, have become central platforms for chip-scale frequency synthesis. Recent breakthroughs in strong-coupling electro-optic modulation further expanded their accessible nonlinear dynamics, unlocking new phenomena and functionalities, but the underlying foundation remains largely unexplored. Here we establish a universal theoretical and experimental framework for nonlinear cavity combs under arbitrary electro-optic modulation by introducing a general evolution equation (GEE) that transcends the mean-field Lugiato–Lefever equation. The GEE reduces to a discrete-time *Integration Hamiltonian* that provides a frequency-domain formalism unifying strong-coupling electro-optic modulation with photonic synthetic dimensions. Together with a band–wave correspondence linking modulation waveforms to synthetic band structures, the formalism enables programmable spectral control. We further show compatibility between Kerr nonlinearity and strong-coupling electro-optic modulation, highlighting their cooperative dynamics. Our work provides a foundational model for strong-coupling electro-optics in nonlinear combs, opening a route toward chip-integrated, microwave-programmable comb sources for metrology, spectroscopy, and emerging photonic technologies.**

## Introduction

Over the past two decades, nonlinear optical frequency combs in micro-resonators have reshaped modern photonics[1], enabling breakthroughs in optical communications[2,3], interconnects[4,5], computing[6–8], ranging[9–11] and precision metrology[12,13]. Among different implementations, Kerr combs have emerged as a leading platform, offering octave-spanning generation in compact, integrated devices[14,15]. Yet, they typically face a trade-off between repetition rate and spectral bandwidth[16], limiting scalability and flexibility. The advancement of the thin-film lithium niobate (TFLN)[17,18] has brought high-speed electro-optic (EO) modulation into this landscape, enabling comb generation with microwave-rate repetition and versatile spectral control. This progress has given rise to cavity EO[19–22] and hybrid EO–Kerr[23–26] combs. However, they are both constrained by the relatively weak EO interaction, with stringent pump-detuning control requirements and limited dynamical richness[16].

The recent emergence of strong-coupling EO modulation offers a promising pathway to overcome these challenges and to open new regimes for EO and hybrid EO–Kerr combs, where the EO-induced coupling strength exceeds the cavity's free spectral range. This regime unlocks a wealth of unexplored phenomena—from robust pump detuning with multi-pulse dynamics to fully programmable spectral shaping[27,28]. Beyond practical advantages, it also enriches the synthetic dimension and non-Hermitian physics[29–31], enabling phenomena such as multi-long-range interactions with a single-tone modulation and synthetic energy-band overlap, which are unattainable in weak-coupling regimes. Yet, despite this promise, previous models, i.e. conventional tight-binding Hamiltonian $\widehat{H}_{tight}$ for EO and Lugiato–Lefever equation (LLE) with modulation term for EO-Kerr, fail to explain either these effects or the interplay between strong-coupling electro-optics and other optical nonlinearities including Kerr effect[30,31]. Developing a unified framework that captures nonlinear comb dynamics under arbitrary EO modulations is therefore crucial—not only to reveal the underlying new physics but also to guide the development of reconfigurable, high-bandwidth, and energy-efficient nonlinear comb sources for the next generation of optical and quantum technologies.

Here we address these challenges by establishing a universal theoretical and experimental framework for nonlinear frequency combs under EO modulation with arbitrary microwave waveforms and modulation strengths, which in particular extends to strong-coupling regimes. Central to this framework is a general evolution equation (GEE) that describes resonators in continuous time beyond the mean-field Lugiato–Lefever equation[32,33], enabling accurate modeling of EO modulation even in the presence of Kerr effects (Fig. 1a). We further show that for pure EO cavities the GEE reduces to a rigorous, discrete-time *Integration Hamiltonian* $\widehat{H}_{ig}$. It transcends conventional tight-binding Hamiltonian and clarifies frequency-domain mode coupling in the emerging synthetic dimension[29]. This extends beyond the recent development of discrete-time formalism that did not provide a well-defined Hamiltonian[27]. To demonstrate both generality and robustness, we experimentally validated the formalism on two complementary platforms: a fiber-cavity system and an integrated thin-film lithium niobate photonic circuit, bridging discrete-component and chip-integrated implementations. Within this framework, we uncover a fundamental band–wave correspondence (BWC), directly linking temporal modulation waveforms $V(t)$ to synthetic band structures $E(k)$, mediated by cavity dynamics in the detuning–angle $(\Delta, \varphi)$ parameter space (Fig. 1b). Leveraging these formalisms, we demonstrate tailored triangular-wave modulation to achieve independent EO comb sideband control, including single-sideband generation.

Finally, incorporating Kerr nonlinearity, we show that GEE offers more precise simulation and modeling than the LLE and reveals the mechanism of soliton band drift, enabling deterministic soliton addressing and EO pulse–Kerr soliton co-excitation, for the first time. In particular, we prove that stronger EO modulation effectively enhances pump power and anomalous dispersion, resolving Kerr combs' repetition-rate–bandwidth trade-off.

## Results

### Introduction of general evolution equation

We first discuss the general evolution equation (GEE) that captures both the strong-coupling EO modulation and Kerr nonlinearity. The complete GEE reads:

$$\frac{da(\varphi,t)}{dt} = \left(-\frac{\kappa}{2} - i\Delta + ig(\varphi,t)\right) a(\varphi,t) + \sqrt{\kappa_e}\, a_{in} t_R \sum_{n\in\mathbb{Z}} \delta\left(t + \frac{\varphi}{\omega_R} - n t_R\right)$$
$$+ i\frac{D_2}{2}\frac{\partial^2}{\partial\varphi^2} a(\varphi,t) + i\sigma |a(\varphi,t)|^2\, a(\varphi,t).$$

The coordinates $(\varphi, t)$ represent the azimuthal angle and time in the co-rotating frame, transformed from the laboratory frame $(\varphi', t')$ via: $\begin{cases} \varphi = \varphi' - \omega_R t' \\ t = t' \end{cases}$. The operator $a(\varphi)$ denotes the annihilation operator of the envelop field at coordinate $\varphi$, and $a(\varphi,t) = \langle a(\varphi) \rangle (t)$ represents its expectation value at time $t$. The total loss rate is given by $\kappa = \kappa_i + \kappa_e$, where $\kappa_i$ and $\kappa_e$ are the internal and external loss rates, respectively. The angular frequencies of the central cavity mode and the pump are denoted by $\omega_0$ and $\omega_p$, with detuning $\Delta = \omega_0 - \omega_p$. Key cavity parameters include the angular repetition frequency $\omega_R$, repetition rate $f_R$ and round-trip period $t_R$, $f_R = 1/t_R = \omega_R/(2\pi)$. The local modulation strength $g(\varphi,t)$ is defined in terms of the local microwave voltage as: $g(\varphi,t) = \pi f_R \frac{V_{loc}(\varphi,t)}{\alpha V_\pi}$, here $\alpha$ is the modulation-region ratio, $V_\pi$ is the half-wave voltage, and $V_{loc}(\varphi,t) = V(t-\varphi/\omega_{\mathrm{R}})$ for a travelling wave electrode, with $V(t)$ representing the applied microwave voltage. The dispersion and Kerr nonlinear coefficients are denoted by $D_2$ and $\sigma$, respectively.

The periodic pump $\sqrt{\kappa_e}\, a_{in} t_R \sum_{n\in\mathbb{Z}} \delta\left(t+\frac{\varphi}{\omega_R}-nt_R\right)$ introduced by GEE breaks the mean-field approximation underlying the LLE and is the key advance toward universal EO-modulated cavity modeling. The cavity repetition rate is intrinsically set by the discrete pump injection period. The LLE replaces this discrete injection with a continuous drive and enforces an asymptotic limit for repetition frequency $\omega_R' \to \infty$, while assuming negligible field variation within one round trip: $\left|\left(a(\varphi,t+t_R)-a(\varphi,t)\right)/a(\varphi,t)\right| \ll 1$. This approximation becomes invalid when the field experiences order-unity change per round trip, as occurs in the strong-coupling regime where the EO coupling rate $g_0 = \max\limits_{\varphi}\left|\int_0^{t_R} f_R g(\varphi,t)\ dt\right|$ becomes comparable to repetition rate $\omega_R$ and thus non-perturbative, leading to mean-field failure in both EO-dominated and EO–Kerr coexisting cavities. In particular, the split-step Fourier transform (SSFT) simulation method of LLE treats pump injection as continuous in time and discretizes it numerically with a time step $\Delta t$, imposing an artificial repetition rate $\omega_R' = 2\pi/\Delta t \neq \omega_R$ and misrepresenting the cavity dynamics. In contrast, GEE-based time evolution restores the natural $t_R$-periodic pump, potentially resolving the long-standing discrepancy between Kerr theory, simulations, and experiments (see SI for the LLE- and GEE-based simulation comparison).

**Formalism of *Integration Hamiltonian* and band-wave correspondence**

The periodic pump of GEE precludes stationary solutions on a continuous time scale. When Kerr nonlinearity is absent and dispersion is negligible[34], the system can be viewed on a discrete time scale, since the intracavity field passes through the EO modulator once per round trip. We therefore develop a simplified frequency-domain coupling-based Hamiltonian formalism for nonlinear systems with pure EO modulation, yielding discrete-time stationary solutions.

Experimental monitoring at the through-port samples the field in the $(\varphi,t)$ parameter space along a trajectory: $\varphi(t) = -\omega_R t$ (equivalent to the fixed monitoring port in the lab frame, $\varphi' = 0$) (Fig. 1c). To predict the field intensity at a given azimuthal angle $\varphi$ sampled at discrete times $T = -\varphi/\omega_R + nt_R (n \in \mathbb{Z})$, we integrate the GEE over a round-trip time $t_R$, to obtain the discrete-time *Integration Hamiltonian* at $\varphi$ :

$$\hat{H}_{ig}^{(\varphi)} = G(\varphi)a(\varphi)^{\dagger}a(\varphi)$$

The evolution coefficient reads $G(\varphi) = if_R\left(e^{-i\Delta t_R}U(\varphi) - 1\right)$, where $U(\varphi) = e^{\int_0^{t_R} ig(\varphi,t)\,dt}$ denotes the round-trip evolution operator of light field at $\varphi$, independent of the detuning frequency $\Delta$. The total Hamiltonian is obtained by spatial integration, $\widehat{H}_{ig} = \int \widehat{H}_{ig}^{(\varphi)} d\varphi/(2\pi)$, and can be further expanded in the spectral domain.

$$\widehat{H}_{ig} = \sum_{\mu}\sum_{m} G_m a_{\mu+m}{}^{\dagger} a_{\mu}$$

Here $\{a_\mu\}$ denote the spectral modes with eigenfrequencies $\{\omega_\mu\}$, where μ represents the relative mode index. The Fourier transform relation yields: $a(\varphi) = \sum_\mu a_\mu\, e^{i\mu\varphi}$, and $G_m$ are the frequency domain coupling coefficients obtained by Fourier transforming $\widehat{H}_{ig,\varphi}$ with respect to $\varphi$. This spectral-domain formalism enables direct calculation of EO comb spectra by solving the corresponding Heisenberg's equations without invoking rotating-wave approximation (RWA) to the pump terms, setting $\hbar = 1$,

$$\frac{\partial a_\mu}{\partial T} = i\left[\widehat{H}_{ig}, a_\mu\right] - \frac{\kappa}{2} a_\mu + \sqrt{\kappa_e} a_{in}\, \delta_{0,\mu} = 0$$

where $\frac{\partial a_\mu}{\partial T} = \frac{a_\mu(T+t_R) - a_\mu(T)}{t_R}$ denotes the discrete-time difference of the field. In contrast, applying the pump-RWA to restrict the pump term to the central mode in the continuous-time spectral-domain equations is mathematically equivalent to the mean-field approximation underlying the LLE—both approximations break down in the strong-coupling regime (see SI for details).

Finally, we apply the rotating-frame transformation, and rotating-wave approximation (RWA) to obtain the conventional Hamiltonian formalism that is widely used for photonic synthetic dimension[18,29,35]. In this case, the *Integration Hamiltonian* can be written equivalently as

$$\widehat{H}_{ig}(t) = \sum_{\mu} \omega_\mu a_\mu{}^{\dagger} a_\mu + G(t) \sum_{\mu}\sum_{m} a_{\mu+m}{}^{\dagger} a_\mu.$$

For a single sinusoidal modulation at zero detuning, $g(\varphi,t) = g_0 cos(\varphi)$, there is $G(t) = if_R\left(e^{ig_0 cos(\omega_R t) t_R} - 1\right)$ which reduces to $G(t) = -g_0 cos(\omega_R t)$ in the weak-coupling case $g_0 \ll \omega_R$.

To verify the *Integration Hamiltonian* formalism, we perform experiments on two complementary platforms (Fig. 2a). We show that the measured strong-coupling EO comb spectra agree well with both the GEE time-evolution results and the stationary solutions obtained from the Heisenberg equations derived from $\widehat{H}_{ig}$ (Fig. 2b). Moreover, we experimentally measure the inter-mode coupling strength $|G_m|$, which governs energy exchange over one cavity round trip (Fig. 2c).

The discrete timescale *Integration Hamiltonian* requires a well-defined energy band. The synthetic-dimension concept draws an analogy between the one-dimensional set of spectral modes $\{a_\mu\}$ and a one-dimensional atomic chain in solid-state physics. In this picture, $\varphi$ plays the role of the quasi-momentum, namely, in the first Brillouin zone $k = \varphi \in [0,2\pi]$. The excitation condition for a mode at azimuthal angle $\varphi$ is $\hat{H}_{ig}^{(\varphi)} = 0$, which defines an effective eigenenergy shift $\omega_0 \to \omega_0(\varphi)$ for each mode, yielding the band structure: $\mathrm{E}(k) = \omega_0(\varphi) = \omega_0 - \pi f_R \frac{V(t=-\varphi/\omega_R)}{V_\pi} - n\omega_R$, where $n$ is the Floquet index (see SI for details). This expression establishes a direct band-wave correspondence (BWC): multiple $\omega_R$ -spaced bands precisely align with the shape of the modulation wave $V(t)$. We experimentally confirm this and observe that overlaps between these bands induce more than two pulses per round trip (Fig. 3).

The *Integration Hamiltonian* formalism goes beyond standard Floquet-Hamiltonian approaches[36]: while both can capture spatial-domain dynamics and band structures, only our framework yields a spectral-domain Hamiltonian and coupled-mode equations without invoking the rotating-wave approximation (RWA), highlighting its importance to cavity electromagnetism.

## Directional control of EO combs enabled by triangular wave modulation

Leveraging the $\hat{H}_{ig}$ and BWC, we engineer EO combs via synthetic band tailoring. We employ triangular waves as a basis, characterized by modulation strength $g_0 = \pi f_R \frac{|V_{max}|}{V_\pi}$ and asymmetry ratio $\gamma = \frac{t_{up}-t_{down}}{t_{up}+t_{down}}$ ($t_{up}$ and $t_{down}$ denote the rising and falling durations) (Fig. 4a).

For symmetric triangular waves ($\gamma = 0$), measured coupling spectra $|G_m| - m$ are symmetric and concentrated at a single dominant order $m_0 = 4\frac{g_0}{\omega_R}$: one dominant coupling-order at a specific modulation strength. This sharply contrasts with the diffuse, multi-order spectrum generated by sinusoidal driving, thereby highlighting the triangular wave's role as a fundamental modulation basis. Stronger modulation increases $m_0$, producing a clear spectral periodicity, as confirmed experimentally (Fig. 4b). For coupling spectra with $\gamma \in [-1,1]$, at extreme $\gamma$, one side exhibits strong low-order coupling, while the other shows only weak high-order coupling. Such directional coupling enhances energy flux in the preferred-coupling direction and produces a more prominent sideband, validated experimentally (Fig. 4c). This sideband asymmetry can also be explained by an energy band picture combining concepts of group velocity and density of states[37] (see SI for details).

We achieve single-sideband control of the EO comb using triangular waves with tailored strength and asymmetry. This arises because EO-comb dynamics are governed by the band slope $k = \partial\omega_0(\varphi_0)/\partial\varphi$ near the excitation energy $\omega_0(\varphi_0)\sim\omega_p$, rather than the global wave. The nearly identical combs generated by a complete triangular wave and its peak-truncated variant (Fig. 4d) confirm the slope-dominant nature. Furthermore, single-sideband control is realized by tuning $g_0$ and $\gamma$ such that one band slope near the excitation is preserved while the other is varied. Measured modulation waves and their spectra (Fig. 4e) verify the effectiveness of this approach.

**Soliton band drift and deterministic Kerr soliton addressing**

When an optical cavity is pumped with high power, Kerr nonlinearity must be incorporated. Based on the GEE, we introduce the concept of soliton band drift to describe dynamics involving multiple nonlinearities (Fig. 5a). The BWC principle remains valid here: in a blue-to-red frequency sweep $\omega_p(t)$, EO modulation distorts the synthetic energy band $\omega_0(\varphi)$, deforms the modulation instability (MI) regime and confines Kerr solitons to the red-detuning region $\omega_0(\varphi) > \omega_p(t)$. A soliton at $\varphi_0$ drifts with velocity determined by the band slope $k = \partial\omega_0(\varphi_0)/\partial\varphi$. With a perturbative Lagrangian approach, we derive the soliton drifting velocity as:

$$v_d = -d_2 k/3$$

where the normalized dispersion reads $d_2 = 2D_2/\kappa$ (see SI). The numerically simulated $v_d$ agree excellently with the theory (Fig. 5b).

Deterministic single-soliton operation is desirable because multi-soliton states exhibit spectral modulation and temporal complexity arising from uncontrolled relative soliton positions. To realize soliton addressing, a single soliton is stabilized at the band minimum $\omega_{0min}$, while others emerging at non-zero band slope should be eliminated. The extra solitons drift toward the band minimum $\omega_{0min}$ at a velocity $v_d \propto -k$, while during the frequency sweep, the MI boundary contracts in the same direction at $v_{MI} \propto -1/k$. Soliton-elimination occurs when solitons collide with the MI boundaries, which requires a large slope $k$ to satisfy $v_d > v_{MI}$ (Fig. 5a).

Since the drifting velocity $v_d \propto -d_2 k$, the soliton-addressing condition follows a complementary scaling law between $d_2$ and $k$ ($k$ is proportional to the modulation strength $g_0$ for a given modulation waveform). We also numerically observe that increasing the pump power $P_{in}$ improves the success probability of single-soliton addressing; the underlying mechanism remains to be theoretically clarified. We quantify the probability of successful soliton addressing (the fraction of simulations that end in a single-soliton state) under EO modulation by scanning ($g_0$, $d_2$) and ($g_0$, $P_{in}$) respectively (Fig. 5c). For both sinusoidal and triangular modulation, stronger EO modulation significantly alleviates the stringent requirements on anomalous dispersion and pump power, thereby enabling the realization of broad-bandwidth and low-repetition-rate (within large cavities) hybrid EO–Kerr combs. Furthermore, the triangular wave leads to a uniform slope $k$ across the band, outperforming the sinusoidal wave in soliton addressing, which in contrast exhibits near-zero $k$ near the band minimum. By extending the parameter space of $(g_0, P_{in})$, we identify distinct dynamical phases for sinusoidal modulation (Fig. 5d) with the corresponding detuning–angle $(\Delta, \varphi)$ phase diagram displayed in Fig. 5e. We especially display the EO pulse-Kerr soliton co-excitation, for the first time.

## Discussion

In summary, we have established a universal framework for nonlinear cavity frequency combs under strong-coupling EO modulation. The general evolution equation (GEE) provides a continuous-time description compatible with Kerr nonlinearities, while the reduced *Integration Hamiltonian* offers a frequency-domain formalism consistent with synthetic dimension approaches for EO-dominated cavities. Together, these models furnish a rigorous foundation for exploring topological and non-Hermitian dynamics[30,38–40] in strong-coupling EO combs. Experimental validation across both fiber-cavity and thin-film lithium niobate platforms confirms the universality of the framework from bulk to chip scale. Another central outcome of this work is the band–wave correspondence (BWC), which directly links modulation waveforms to synthetic band structures. This principle opens an entirely new degree of freedom for synthetic band tailoring. Leveraging *Integration Hamiltonian* and BWC in combination, we demonstrate directional control of EO comb mode coupling with asymmetric triangular-wave modulation, enabling programmable spectrum shaping and single-sideband comb generation. Such asymmetric optical signal processing holds immediate promise for applications in sensing[41], FMCW LiDAR[42–44], optical communications[45–48], and dual-comb spectroscopy[49–52].

Beyond EO systems alone, the compatibility of the GEE with Kerr nonlinearity unlocks the possibility of strong-coupling EO–Kerr combs, which could combine the complementary advantages of EO precision and Kerr broadband solitons. Thin-film lithium niobate platforms[17,18]—already capable of realizing both EO[20,21] and Kerr combs[53–55]—provide a viable path toward achieving such hybrid systems experimentally. Future studies are expected to extend this approach to encompass more complex $\chi^{(2)}$ nonlinear effects—such as second-harmonic generation, optical parametric oscillation, and spontaneous parametric down-conversion—along with cascaded difference- and sum-frequency generation.

Our results therefore establish the first versatile and unified framework for strong-coupling EO combs, capturing universal dynamics in the presence of multiple nonlinearities. This framework may serve as a cornerstone for next-generation comb science and technology, enabling chip-integrated, microwave-programmable frequency comb sources with transformative opportunities for precision metrology[12,13], adaptive spectroscopy[49,56], optical communications[2,3,57], and emerging photonic computing[6–8,58].

## Materials and methods

Supplementary Information is available for this paper, providing details of the theoretical derivations, experimental platforms, and numerical simulations presented in our article.

**Acknowledgements:** Y.X. acknowledges Qixuan Zhou, Cheng Wang, Hanfei Hou, Tong Ge, Junting Bie for helpful discussion. Y.X. acknowledges Hang Yuan, Fan Yang for administrative support, as well as Tom Chen (Flexcompute).

**Author contributions:** Y.X., T.L., and Y.H. conceived the project and contributed to theoretical modeling. X.L. constructed the fiber cavity. G.W. fabricated the TFLN chip. Y.X. and T.L. performed the measurement. Y.X. performed the numerical simulations, and data processing. Y.X. and Y.H. wrote the manuscript with input from all authors. C.C., Y.L. and M.W. helped the project. Y.H., D.Z., Q.G. and Y.L. supervised the project.

**Data availability:** The data that support the findings of this study are available from the corresponding author upon reasonable request.

**Code availability:** The code used to produce the numerical simulations within this paper is available from the corresponding authors upon reasonable request.

**Funding:** This research is supported by the National Key Research and Development Program (2023YFB3211200), Quantum Science and Technology-National Science and Technology Major Project (2024ZD0302500, 2024ZD0302501, 2024ZD0301600), National Science Foundation of China (NSFC-12474321), National Research Foundation Singapore (NRF-NRFF15-2023-0005).

**Competing interests:** The authors declare no competing interests.

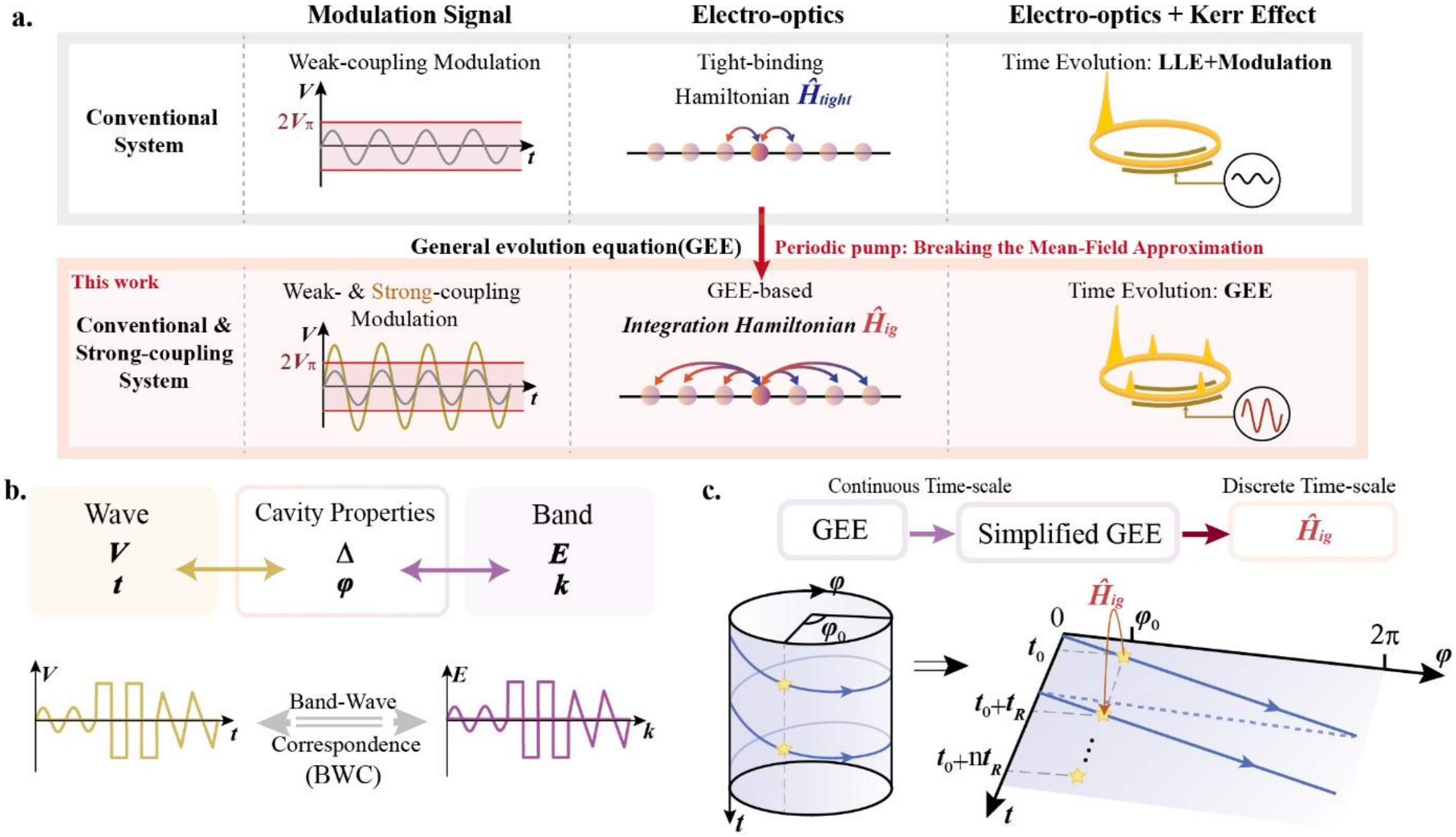

**Figure 1 | Nonlinear system under strong-coupling electro-optic (EO) modulation with general evolution equation (GEE) and band-wave correspondence (BWC). a,** GEE breaks the mean-field approximation by introducing the periodic pump, allowing a unified framework for EO and EO–Kerr system in both weak- and strong-coupling regime. **b,** The synthetic dimensional band structure ($E - k$) corresponds to the modulation wave ($V - t$), linked through resonator dynamics in the parameter space of detuning-angle ($\Delta - \varphi$). **c,** In EO-dominated, low-power cavities, the GEE reduces to an *Integration Hamiltonian* which captures stationary solutions on a discrete time scale. In optical cavities, the angle $\varphi$ and time $t$ (co-rotating frame) define a cylindrical parameter space. Experimental monitoring samples the space along the trajectory $\varphi(t) = -\omega_R t$ (blue line). For an azimuthal angle $\varphi_0$ sampled at discrete times $t_0 + nt_R, n \in \mathbb{Z}$, the discrete-time Heisenberg equation of motion defines the *Integration Hamiltonian* $\hat{H}_{ig}$.

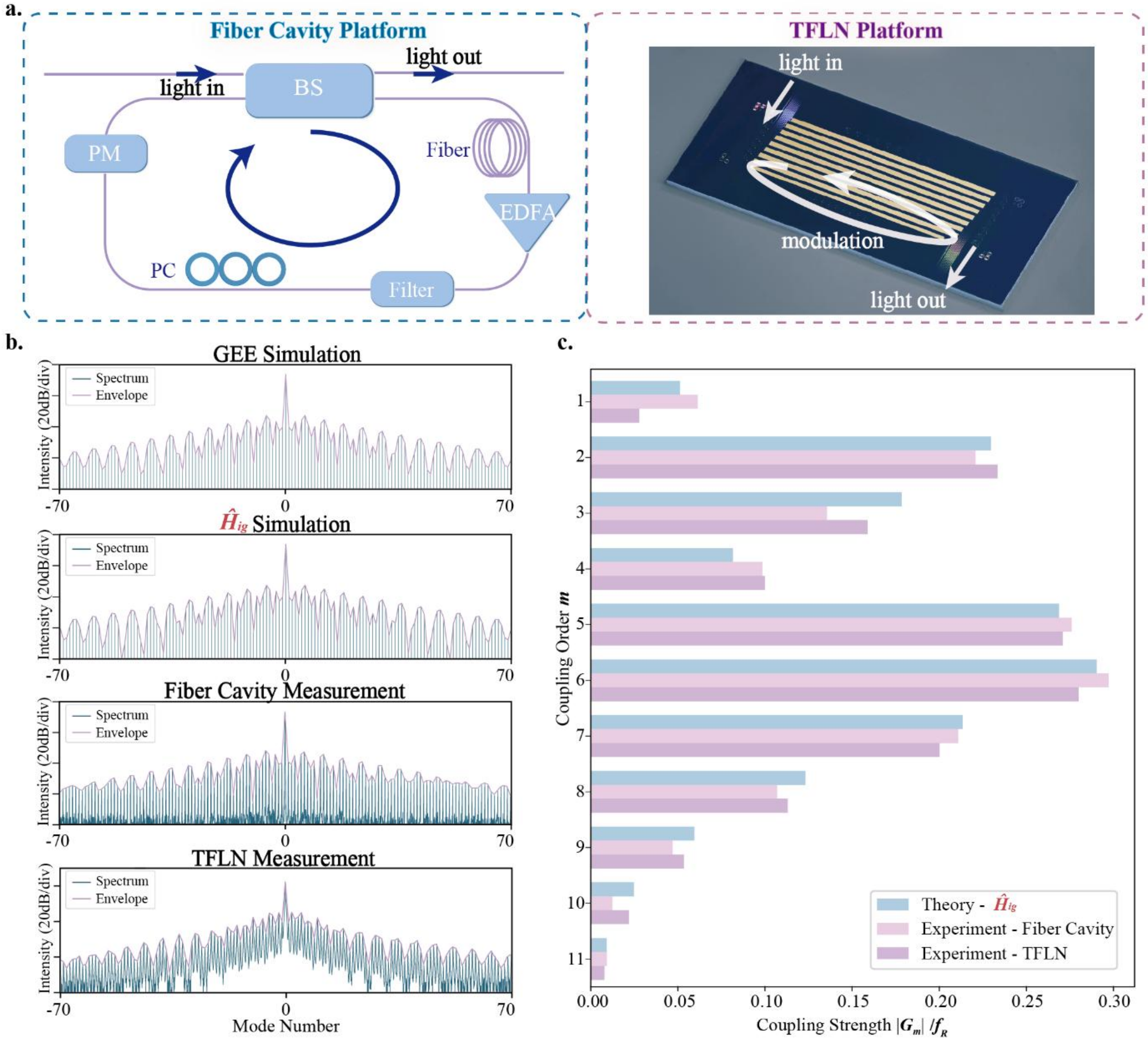

**Figure 2 | Validation of *Integration Hamiltonian* formalism. a,** Illustration of the experimental platforms. Left: fiber cavity platform; PC, polarization controller; PM, phase modulator; BS, beam splitter; EDFA, erbium doped fiber amplifier. Right: Thin-film lithium niobate (TFLN) photonic chip, with fiber-to-fiber coupling configuration. **b,** Simulated and measured spectra under a microwave drive $|V_{max}| = 2.3V_\pi$ ($g_0 = 1.15\omega_R$). **c,** The measured and calculated coupling strengths $|G_m| - m$ for the *Integration Hamiltonian*.

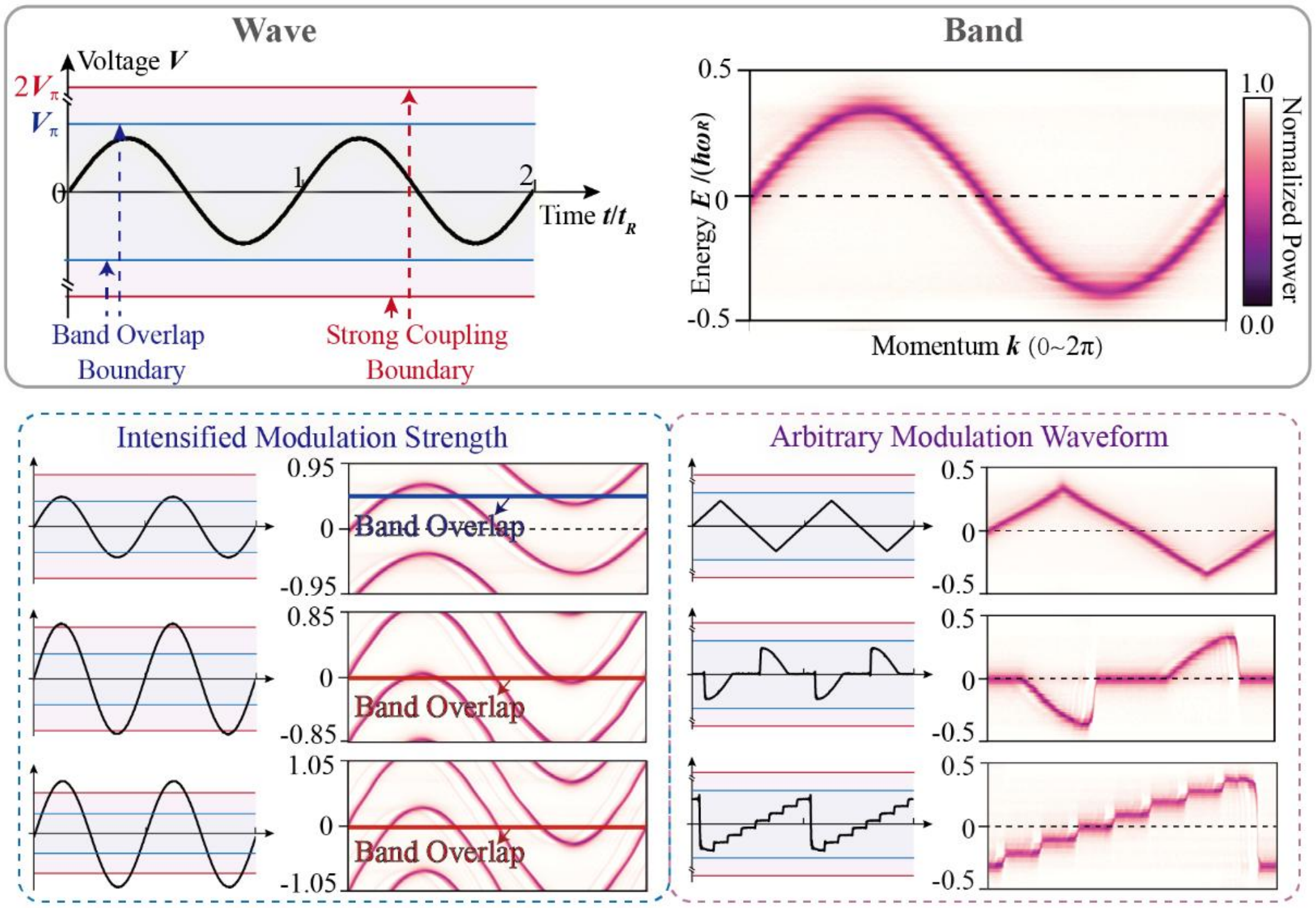

**Figure 3 | Experimental demonstration of band-wave correspondence (BWC).** The measured modulation waveform and corresponding band structure in a fiber cavity, validating the BWC principle at arbitrary modulation strength and waveform. With increasing modulation strength, beyond the band-overlap boundary $|V_{max}| > V_\pi$ ($g_0 > 0.5\omega_R$), band overlaps occur at detuning $\Delta \neq 0$; beyond the strong-coupling boundary $|V_{max}| > 2V_\pi$ ($g_0 > \omega_R$), band overlaps occur at $\Delta = 0$.

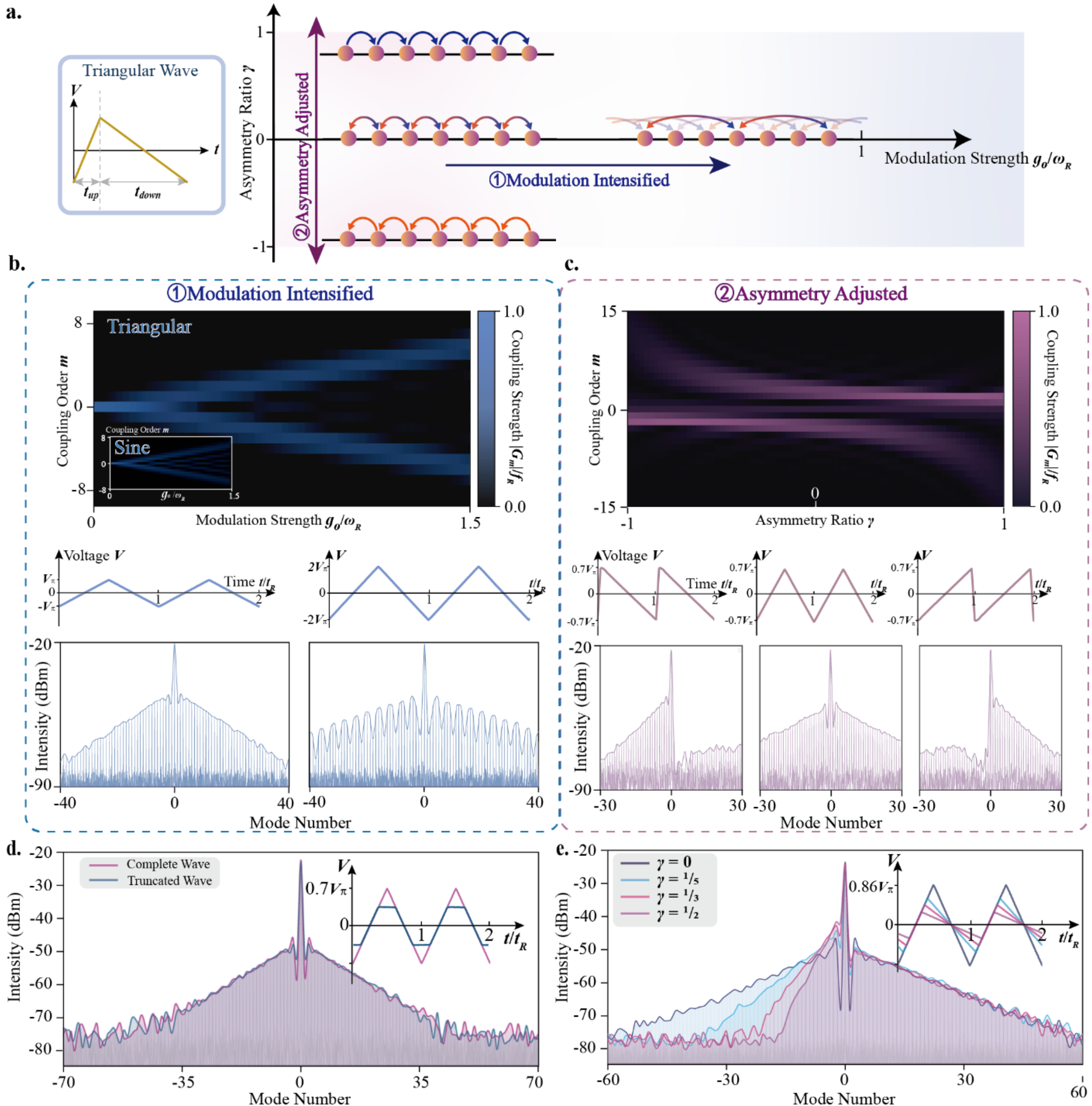

**Figure 4 | EO combs: Triangular waveform as sideband engineering enabler. a,** The triangular modulation, characterized by modulation strength $g_0$ and asymmetry $\gamma$ enables engineering of EO comb frequency-domain mode coupling. **b,** Coupling strength spectrum $|\mathrm{G_m}| - \mathrm{m}$ as a function of $g_0$. Inset: corresponding spectrum under sinusoidal wave modulation. Experiments confirm that increasing $g_0$ elevates the coupling order in EO comb spectra. **c,** Coupling strength spectrum $|\mathrm{G}_m| - m$ as a function of $\gamma$. Experiments confirm that this directional coupling enhances the sideband in the strongly coupled direction. **d,** Experiments confirm that EO comb dynamics are governed by band slope near the excitation energy $\omega_p$. **e,** Experimental demonstration of unidirectional EO comb control with strength-asymmetry-tailored triangular wave.

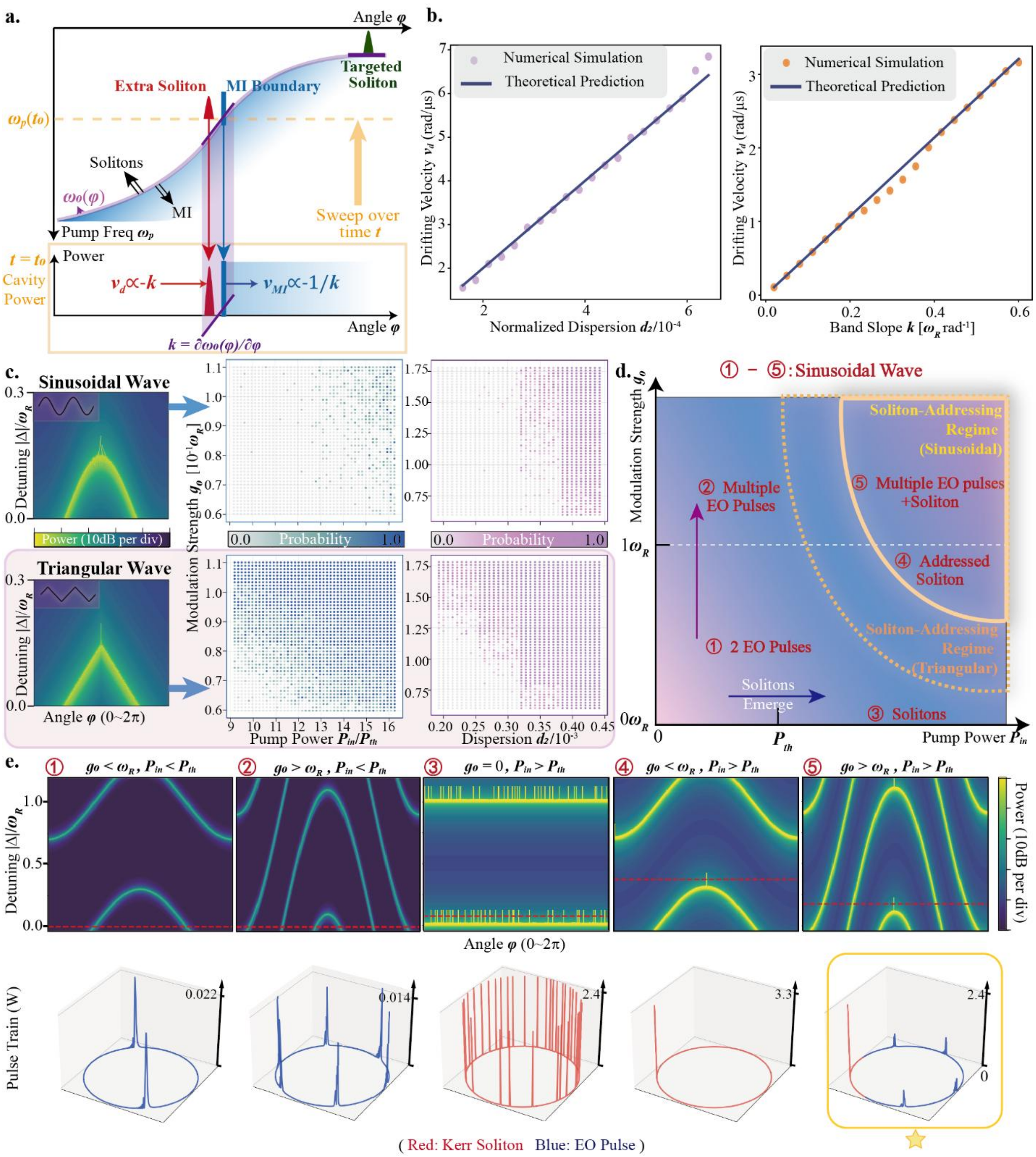

**Figure 5 | Soliton band drift theory and strong-coupling EO–Kerr phenomena. a,** In a blue-to-red frequency sweep, the parameter space of pump frequency $\omega_p$ and angle $\varphi$ separates into soliton-existence and modulation-instability (MI) regimes, separated by a boundary parallel to the synthetic band $\omega_0(\varphi)$. Soliton addressing occurs when extra solitons generated at non-zero band slope $k$ drift toward and collide with the contracting MI boundary, which favors larger band slopes $k$ and stronger modulation strengths $g_0$. **b,** Numerical simulations of drifting velocity $v_d$ versus the band slope $k$ and the normalized dispersion $d_2$. **c,** Based on sinusoidal and triangular wave modulation, the probability of successful soliton addressing is quantified by scanning ($g_0$, $P_{in}$) and ($g_0$, $d_2$) ($P_{in}$ is the

pump power; $P_{th}$ is the threshold power of parametric oscillation). **d,** Phase diagram of cavity states under sinusoidal wave modulation in the parameter space of $(g_0, P_{in})$. **e,** Representative $(\Delta, \varphi)$ phase diagrams and corresponding pulse-train dynamics, evaluated at the detunings marked by red horizontal lines. We especially display ⑤: the EO pulse-Kerr soliton co-excitation, for the first time.